\documentclass[10pt,journal,compsoc]{IEEEtran}
\newif\ifpeerreview

\peerreviewfalse

\usepackage[nocompress]{cite}
\usepackage{url}
\usepackage{amsmath,amssymb,graphicx}
\usepackage{tikz}
\usepackage{pgfplots}
\usepackage{xcolor}
\usepackage{hyperref}
\usepackage{gensymb}

\DeclareMathOperator*{\argmin}{arg\,min}

\usepackage{lipsum} %

\usepackage[switch]{lineno}

\newcommand{\paperID}{0072}

\title{A Gaussian Parameterization for Direct Atomic Structure Identification in Electron Tomography}

\author{Nalini~M.~Singh,
        Tiffany~Chien,
        Arthur~R.C.~McCray,
        Colin~Ophus,    
        and~Laura~Waller%
\IEEEcompsocitemizethanks{\IEEEcompsocthanksitem N. Singh, T. Chien, and L. Waller are with the Department of Electrical Engineering and Computer Science, University of California, Berkeley, CA 94709. \protect\\
E-mail: see http://www.nalinimsingh.com
\IEEEcompsocthanksitem A. McCray and C. Ophus are with the Department of Materials Science, Stanford University, Palo Alto, CA 94305.}%
}

\begin{document}

\IEEEtitleabstractindextext{%
\begin{abstract}
Atomic electron tomography (AET) enables the determination of 3D atomic structures by acquiring a sequence of 2D tomographic projection measurements of a particle and then computationally solving for its underlying 3D representation. Classical tomography algorithms solve for an intermediate volumetric representation that is post-processed into the atomic structure of interest. In this paper, we reformulate the tomographic inverse problem to solve directly for the locations and properties of individual atoms. We parameterize an atomic structure as a collection of Gaussians, whose positions and properties are learnable. This representation imparts a strong physical prior on the learned structure, which we show yields improved robustness to real-world imaging artifacts. Simulated experiments and a proof-of-concept result on experimentally-acquired data confirm our method's potential for practical applications in materials characterization and analysis with Transmission Electron Microscopy (TEM). Our code is available at \url{https://github.com/nalinimsingh/gaussian-atoms}.
\end{abstract}

\begin{IEEEkeywords} %
Atomic electron tomography, Gaussian splatting
\end{IEEEkeywords}
}

\maketitle

\IEEEraisesectionheading{
  \section{Introduction}\label{sec:introduction}
}
\IEEEPARstart{T}{he} reliable, precise determination of 3D atomic structures of nanomaterials is a fundamental goal of materials science. Knowledge of these 3D structures enables simulation and characterization of material properties from first principles, for use in applications ranging from chemical process engineering to protein design. Transmission electron microscopy (TEM), which uses an electron beam transmitted through a sample of interest, offers resolution at the scale of individual atoms due to the small wavelength of the electron~\cite{williams2009transmission}. Atomic electron tomography (AET) in particular enables the determination of 3D atomic structures by acquiring a sequence of 2D tomographic projection measurements of a particular particle and then computationally solving for its underlying 3D structure~\cite{miao2016atomic}. 

Traditionally, solving for 3D atomic structure from AET projection measurements is a two-step procedure. First, an algorithm reconstructs a volumetric image of a particle of interest. Then, a procedure called \textit{atom tracing} extracts the locations and chemical species of individual atoms within the volume via a combination of algorithmic peak finding and manual post-processing. Although these methods have been used successfully to identify structures of interest~\cite{pelz2023solving,yang2017deciphering,xu2015three}, this approach suffers from several limitations. First, practical imaging constraints limit the quality of the reconstructed volume. For example, physical limits on the set of feasible tilt angles make the tomography inverse problem ill-posed, yielding physically-unrealistic streak-like artifacts in the reconstruction that are fully consistent with the acquired data but do not preserve the known atomic nature of the underlying structure. Second, the atom-tracing step is error-prone and requires substantial manual human labor to accurately identify structures that may contain several thousands of atoms. Since this step is downstream of the image reconstruction stage, image artifacts induced at that stage may throw off peak-finding algorithms.

\begin{figure}[!t]
\centering
\includegraphics[width=\columnwidth]{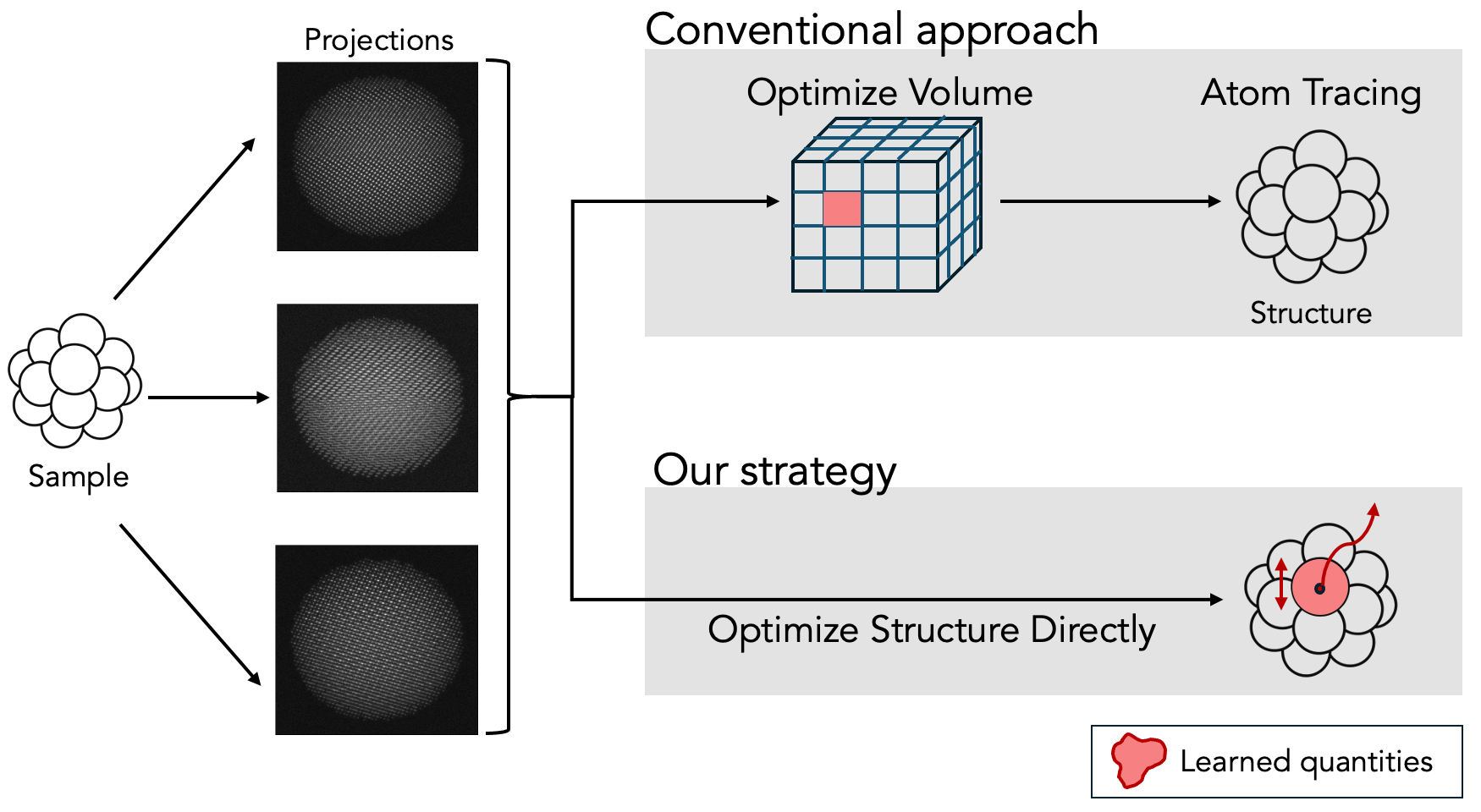}\caption{Traditional methods for atomic electron tomography (AET) first computationally reconstruct a volumetric image of the sample and then identify the underlying atomic structure via a combination of algorithmic peak finding and manual post-processing. Our proposed method simplifies this workflow to solve for the 3D properties of each atom directly. }
\end{figure}

In this paper, we reformulate the tomographic inverse problem to solve directly for the locations and properties of individual atoms, instead of solving for an intermediate voxelized representation that is post-processed into an atomic structure. We model each sample of interest as a collection of Gaussians with learnable locations, sizes, and amplitudes, and we directly optimize these parameters from AET projection measurements. This parameterization imparts a strong physical prior on the associated reconstruction, because in practice atomic electron densities appear Gaussian. While a voxelized representation is flexible enough to represent physically-unrealistic structures, our parameterization can only represent collections of atomic objects, even in the presence of imaging corruptions that would otherwise introduce artifacts. Further, the parameterization makes it easy to enforce physical atom-specific constraints, such as realistic atomic size limits and minimum bond lengths, that are much harder to express in a voxelized representation. Because the learned Gaussian locations directly represent the atomic structure we would like to solve for, our approach simplifies the two-step process from the traditional method and allows us to directly optimize for the structure of interest. We show in simulation and experiment that our parameterization enables fast, accurate atomic structure determination more robust to realistic imaging artifacts than conventional methods.

To summarize, our key contributions are as follows: 
\begin{itemize}
    \item We introduce a novel workflow for atomic structure identification that directly optimizes atomic positions and characteristics.
    \item We introduce a set of physics-based priors that enforce a one-to-one correspondence between atoms and Gaussians and impose meaningful constraints on the reconstructed structure.
    \item In simulation, we demonstrate improvements in reconstruction quality and atomic structure identification compared to traditional methods, especially in the presence of imaging artifacts.
    \item We demonstrate a proof of concept that our method is able to identify atomic structures from real-world experimental projection data.
\end{itemize}

Our proposed framework is a step toward a more streamlined, direct atomic structure identification process that could enable the reconstruction of nanoparticles and experimental datasets that were previously unusable.

\section{Background and Related Work}
In TEM, a beam of electrons passes through a thin sample of interest. Individual electrons scatter as they interact with atoms in the sample and then the detector records the image intensity for processing and visualization~\cite{williams2009transmission}. Scanning transmission electron microscopy (STEM), which we use here, focuses the electron beam with subatomic precision and then raster scans it across the sample to form a 2D image~\cite{pennycook2011scanning}. This 2D image represents a \textit{projection} that integrates the atomic potential in the direction of the electron beam through the sample. Atomic electron tomography (AET) tilts the sample to collect a series of these 2D projections from different angles~\cite{chen2013three}. The 3D atomic structure is then reconstructed computationally by an optimization problem searching for an atomic configuration that is consistent with the whole set of projection measurements.

\subsection{Forward Model}
The relationship between the unknown 3D structure and the observed projections follows the forward model given by:

\begin{equation}
\label{eq:forward_model}
p(\rho, \hat{n}) = \int f(\mathbf{r}) \, \delta(\hat{n} \cdot \mathbf{r} - \rho) \, d\ell,
\end{equation}
\noindent where  
\( \mathbf{r} \in \mathbb{R}^3 \) is the position vector,  
\( \hat{n} \in S^2 \) is the unit direction vector of the projection (where \( S^2 \) is the unit sphere in \( \mathbb{R}^3 \)),  
\( \rho \in \mathbb{R} \) is the perpendicular distance from the origin to the projection plane,  
\( p: \mathbb{R} \times S^2 \to \mathbb{R} \) is the projection data,  
\( f: \mathbb{R}^3 \to \mathbb{R} \) is the function representing the object being imaged,  
\( \delta: \mathbb{R} \to \mathbb{R} \) is the Dirac delta function,  
and \( d\ell \) is the differential line element along the projection path.

If we assume that $f$ is reconstructed on a finite, discrete grid of pixels, we can write the set of equations of the form above as a linear matrix system of equations:
\begin{equation}
\label{eq:mat_forward_model}
\mathbf{p} = \mathbf{A}\mathbf{f} + \boldsymbol{\epsilon},
\end{equation}
where $\mathbf{p} \in \mathbb{R}^N$ combines all acquired projection measurements, $\mathbf{f} \in \mathbb{R}^M$ represents the vectorized 3D object, $\mathbf{A}$ is the system matrix that models the contribution of each voxel in $\mathbf{f}$ to each measurement in $\mathbf{p}$, $M$ is the total number of voxels in the object representation, and $N$ is the number of acquired projection measurements. Here we have also included a measurement noise term $\boldsymbol{\epsilon}\in \mathbb{R}^N$.

\begin{figure}[htbp]
\centering
\includegraphics[width=\columnwidth]{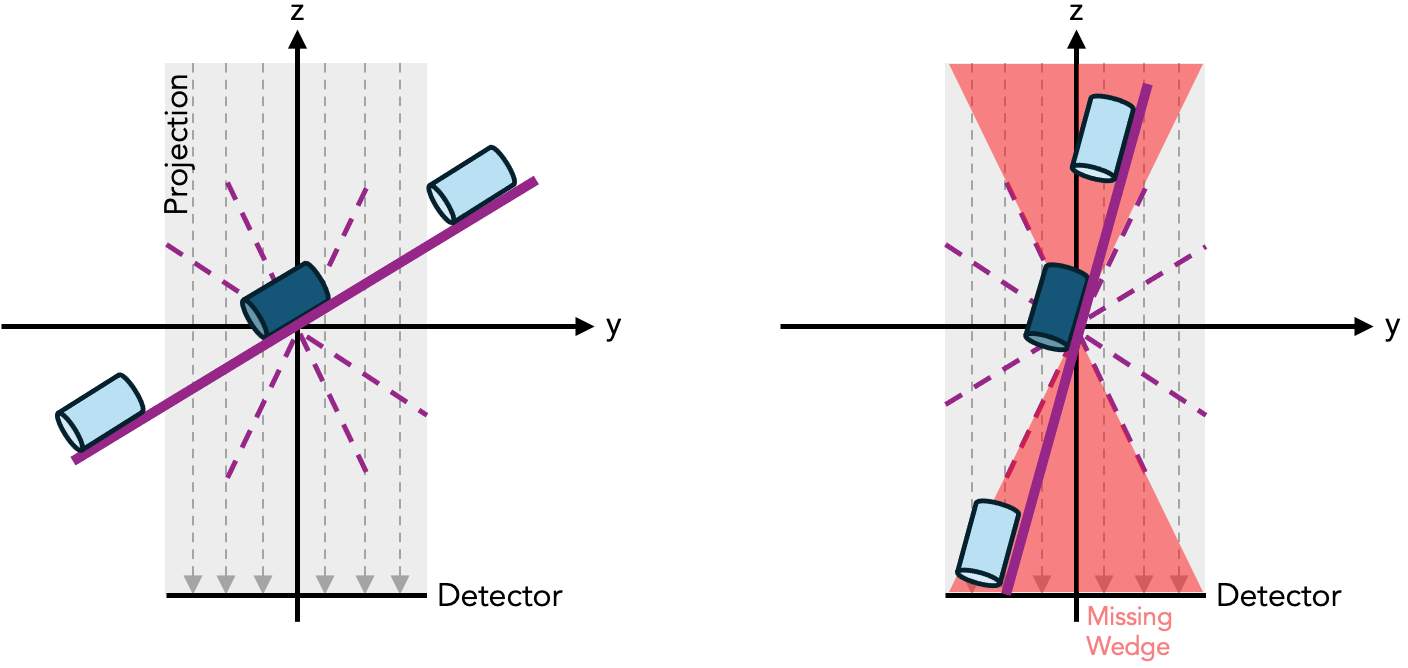}
\caption{Tomographic acquisition setup showing the sample that will be reconstructed (dark blue) tilted at various angles (dotted purple). Several particles outside of the region of interest (light blue) lie on the sample holder, and at high tilt angles (right), interfere with the acquired projection measurements. This results in the \textit{missing wedge} problem where projection measurements cannot be acquired at certain angles, yielding an underdetermined inverse problem. Even beyond the missing wedge, angles maybe sampled sparsely for dose reduction.}
\label{fig:tomography_setup}
\end{figure}

In the ideal setting, the number of unknowns $M$ is equal to the number of acquired projections measurements $N$ -- i.e., $A$ is a square matrix. In practice, however, most tomographic acquisitions suffer from the ``missing wedge" problem, in which some projection angles are inaccessible due to physical constraints. In AET, as shown in Fig.~\ref{fig:tomography_setup}, the sample and its holder occlude the measurement at high projection angles, making it impossible to tilt the sample through a full 180$^{\circ}$ range. Instead, projections are limited to roughly $\pm70^{\circ}$ from the horizontal axis. The Fourier slice theorem~\cite{bracewell1956strip} states that the 2D Fourier transform of a 2D projection of a 3D volume is equivalent to a slice of the object's 3D Fourier transform. Thus, we can think about the missing projection measurements as lost information about the object's Fourier transform in a particular angular direction. Mathematically, this means $\textbf{A}$ is a wide matrix with more columns than rows, leading to an underdetermined inverse problem. 

Further, to limit the electron dose applied to the sample, angles may be sampled sparsely even in the acquisition region outside of the missing wedge, corresponding to additional lost slices in the object's Fourier space representation and making the inverse problem even less well-determined.

\subsection{Inverse Problem}
Image reconstruction involves solving an inverse problem to recover the 3D atomic potential $\textbf{f}$ from the 2D projections $\textbf{p}$. For a voxelized representation of $f$, the inverse problem is typically formulated as a minimization problem~\cite{vogel2002computational}:
\begin{equation}
\label{eq:voxel_objective}
\hat{\mathbf{f}} = \arg\min_{\mathbf{f}} \left\{ \frac{1}{2} \|\mathbf{A}\mathbf{f} - \mathbf{p}\|_2^2 + \lambda R(\mathbf{f}) \right\}.
\end{equation}
The first term is a \textit{data consistency} term measuring how well the reconstructed atomic potential agrees with the acquired measurements under the tomography forward model. If $\textbf{A}$ is a well-conditioned matrix and the noise $\boldsymbol{\epsilon}$ is not overwhelming, achieving a low data consistency loss is sufficient for solving the inverse problem. However, if $\mathbf{A}$ is poorly conditioned, as in the case of a significant missing wedge, many solutions achieve equally low loss under the data consistency term. These solutions are disambiguated by the second term that applies a regularizer $R(\textbf{f})$ to describe prior knowledge about the atomic potential. For example, atomic images often benefit from regularization with the $\ell_1$ norm for sparsity, $R(\mathbf{f}) = \|\mathbf{f}\|_1$, or the total variation norm, $R(\mathbf{f}) = \|\nabla \mathbf{f}\|_1$~\cite{ren2020multiple}. The relative importance of the data consistency term and the regularizer are traded off by the hyperparameter $\lambda$.

\subsection{Previous Approaches}
Previous approaches~\cite{pelz2023solving,yang2017deciphering,xu2015three} first construct a volumetric reconstruction of the 3D atomic potential and then run a procedure called atom tracing to identify individual atom locations. Both of these steps are discussed below.

\subsubsection{3D Reconstruction}
Filtered backprojection (FBP) is a standard approach for solving the tomography reconstruction problem, where the forward model in Eq.~\ref{eq:mat_forward_model} is inverted by effectively smearing each projection measurement across the voxels that contribute to it. Each projection is high-pass filtered before this process, to account for the less dense sampling of the high-frequency content when filling in Fourier space with projections. 

Unfortunately, this filtering also increases the algorithm's sensitivity to noise and other artifacts. A separate class of iterative techniques, including the simultaneous iterative reconstruction technique (SIRT) and the simultaneous algebraic reconstruction technique (SART)~\cite{andersen1984simultaneous}, start with an initial solution to the problem in Eq.~\ref{eq:mat_forward_model} and refine this solution based on knowledge of the system matrix $\mathbf{A}$. A newer subset of iterative approaches including Equally Sloped Tomography (EST) ~\cite{miao2005equally} and the Generalized Fourier Iterative Reconstruction (GENFIRE) ~\cite{pryor2017genfire} additionally operate in Fourier space to improve these reconstructions. 

None of the previously described approaches incorporate prior information about the atomic potential, i.e. the regularizer term in Eq.~\ref{eq:voxel_objective}. To that end, a different class of Model-Based Image Reconstruction (MBIR) approaches directly optimize Eq.~\ref{eq:voxel_objective}, including such a prior~\cite{venkatakrishnan2014model,van2009model}.

More recently, deep learning has become a popular tool for solving the tomographic inverse problem. Implicit neural representations (INRs) use a multi-layer perceptron to map input coordinates $(x,y,z)$ to a function value in continuous space. In the context of AET, these networks can be used to represent the atomic potential $\textbf{f}$~\cite{chien2023space}. Unlike other deep learning approaches that require an auxiliary dataset for training, INR weights are learned by optimizing the objective function in Eq.~\ref{eq:voxel_objective}, a fully self-supervised loss function. Nearby inputs to the network typically produce similar outputs, imparting an implicit spatial smoothness prior on the learned volume. 

Supervised deep learning methods train convolutional neural networks (CNNs) to map projection data that is corrupted by artifacts (e.g. the missing wedge artifact) to high-quality 3D reconstructions. These approaches typically take advantage of large collections of simulated data to encourage a dataset-specific prior on the output reconstruction. They have shown promise in reducing artifacts and improving resolution~\cite{lee2021single}, but they require a significant amount of ground-truth data for training, which are often difficult to create or evaluate for novel materials. For this reason, in this paper we focus on the setting where no additional data are available beyond the measurements to be reconstructed.

\subsubsection{Atom Tracing}
After a 3D volume is reconstructed, conventional AET methods process the volume to identify the locations of atoms in a process called \textit{atom tracing}. A typical atom tracing strategy~\cite{xu2015three,yang2017deciphering,pelz2023solving} first identifies local maxima from a volume reconstruction. Then, beginning at the peak with the highest intensity, a 3D Gaussian function is fit to the peak. If the distance between the peak and any previously selected peak is larger than a specified threshold (often based on the covalent bond length of the known constituent atoms), the current peak is added to a list of candidate atoms~\cite{yang2017deciphering}. 
Template atoms based on the Gaussians are learned for atomic species present in the particle as well as a "non-atom" category, by averaging the set of fitted Gaussians that are estimated to clearly be part of each category based on histogram intensity. Each candidate atom is labeled as belonging to the category whose template is the closest. This process is repeated iteratively, successively updating the categories and the atom estimates. The learned peaks are manually inspected and reassigned as appropriate.

Artifacts that affect the reconstruction can cause problems at several points in this process. For example, streaking artifacts in the reconstruction caused by the missing wedge may affect the computed local maxima, and/or may cause a poor match between any individual peak and the correct corresponding template atom. This process thus relies heavily on manual inspection, which is a laborious and time-consuming process for 3D particles containing thousands of atoms.

\subsection{Gaussian Splatting}
Our proposed representation, a collection of Gaussians, draws heavily on Gaussian splatting~\cite{kerbl20233d}, a technique from computer vision originally developed for the problem of scene rendering. The idea is to represent \textit{any} photographic scene using a collection of Gaussians, whose positions, covariances, colors, and opacities are learned such that scenes rendered via the rendering equation match photographs taken from multiple views surrounding the subject. Complex non-spherical shapes can be accurately modeled using large numbers of Gaussians with highly anisotropic covariances. The contribution of each Gaussian to a particular scene projection can be computed quickly because the 2D projection of a 3D Gaussian is always a 2D Gaussian. Then, a particular scene projection can be computed by iterating over the collection of Gaussians and integrating the projection components. This process is much more computationally efficient than standard ray tracing, where densities must be integrated over rays covering the entire imaging volume. 

A key technical challenge in Gaussian splatting is appropriately guiding the optimization of the Gaussians, because the number of Gaussians required is unknown a priori. In practice, a data consistency loss based on the scene rendering equation is used to optimize the Gaussian parameters, just as it would be used to optimize voxel values. As the optimization progresses, it progressively adapts the number of Gaussians. Extraneous Gaussians are culled when their opacity drops below a threshold. Gaussians that are insufficient to represent a local shape are replaced with two Gaussians that are further optimized according to the standard process; these Gaussians are detected when they have a positional gradient above some threshold.

Our work uses the same core parameterization and optimization strategy as the original computer vision-based Gaussian splatting approach, but the AET problem differs from scene rendering in several important ways. First, our forward model is different. Unlike camera models, tomographic projections are orthographic, with all measurements on the detector derived from parallel rays, and there is no opacity saturation as in the computer vision model. Further, the goal in the scene rendering setting is novel view synthesis\textemdash views of the scene rendered from new positions should be accurate\textemdash  with no regard to the internal structure being reconstructed. In contrast, the goal of tomography is explicitly to solve for the internal structure accurately. 

These issues have previously been explored in work that adapts Gaussian splatting for computed tomography, where tomographic X-rays are used to produce volumetric images, often of the human body~\cite{nikolakakis2024gaspct,cai2024radiative,zha2024r}. However, unlike the general tomography case, for our application of atomic structure identification, it is not sufficient to accurately estimate the atomic potential with any number of Gaussians. To achieve precise structure determination, we must do so in a way that preserves a 1-to-1 correspondence between Gaussians and atoms in the structure.

\section{Proposed Method}
In this section, we first outline our reformulation of the tomographic inverse problem as a problem to solve for \textit{atoms}, instead of a volume. We then describe our optimization procedure and introduce two physics-based priors that guide the optimization and enable us to achieve 1-to-1 correspondence between learned Gaussians and atoms, thereby solving an atomic structure. Detailed information about our implementation and compute requirements are available in the Supplementary Material. Our code is available at \url{https://github.com/nalinimsingh/gaussian-atoms}.

\subsection{Model}

In our framework, the particle of interest is represented as a collection of atoms. The electrostatic potential for a single atom can be modeled as a Gaussian:
\begin{equation}
f_a(\mathbf{r};\mathbf{r}_a,\Sigma_a,q_a) = q_a \cdot \exp\left(-\frac{1}{2}(\mathbf{r} - \mathbf{r}_a)^T\Sigma_a^{-1}(\mathbf{r} - \mathbf{r}_a)\right),
\end{equation}
\noindent where $\mathbf{r}_a$ is the position of atom $a$, $q_a$ is the amplitude, and $\Sigma_a$ is the covariance of the Gaussian.
For a collection of atoms, the total potential is then
\begin{equation}
f(\mathbf{r}) = \sum_{a=1}^{K} f_a(\mathbf{r}),
\end{equation}
where $K$ is the total number of atoms.
The projection measurements can then be computed according to Eq.~\ref{eq:mat_forward_model}. 

Our inverse problem estimates atomic positions $\{\mathbf{r}_a\}$, variances $\{\Sigma_a\}$, and amplitudes $\{q_a\}$. We do not know a priori how many atoms are in the nanoparticle of interest, so we must also optimize $K$. This is formulated as:
\begin{equation}
 \argmin_{K, \mathbf{r}_a, q_a, \sigma_a} \left\{ \mathcal{L}\left [ \mathbf{p},  \tilde{\mathbf{p}}\left (\sum_{a=1}^{K} f(\cdot; \mathbf{r}_a, q_a, \sigma_a)\right ) \right ]\right \},
\end{equation}
\noindent where $\tilde{\mathbf{p}}$ represents the predicted projection data for a given set of atomic parameters according to Eq.~\ref{eq:forward_model} and $\mathcal{L}$ is a loss function. Qualitatively, this optimization is quite different from the one in Eq.~\ref{eq:voxel_objective}. Of particular note, the unknown nature of $K$ makes the number of optimization variables dynamic. 

The key advantage of our approach is that the final atomic positions are directly available from the optimized Gaussian centers $\{\boldsymbol{r}_a\}_{a=1}^{K}$. This eliminates the need for the additional atom tracing step required in traditional methods. Atoms of different elements exhibit distinct scattering potentials, which are reflected in the optimized $q_a$ and $\Sigma_a$ values and can be leveraged for element identification. 

\subsection{Optimization}
We employ a gradient-based optimization approach to refine the Gaussian parameters based on the consistency between simulated projections and experimental measurements. Our approach to this builds on the optimization scheme proposed in Gaussian splatting~\cite{kerbl20233d}. We initialize Gaussians at random locations across the volume to be reconstructed, and then we optimize according to the loss function. We use an L1 loss between simulated and experimental projections as in traditional Gaussian splatting but remove the SSIM term which we empirically find performs poorly on atomic projection measurements and often encourages Gaussians to overfit small nuisance details. 

 As in traditional Gaussian splatting, the simple analytical expression describing Gaussian densities admits an explicit analytical expression for gradient descent updates to the Gaussian parameters. Instead of using auto-differentiation with respect to the parameters of the Gaussian, they are optimized via a closed-form update. We derive the analogous update for the tomographic forward model and use the corresponding analytical expressions to perform gradient descent, as in prior work on Gaussian splatting for tomography~\cite{zha2024r}.

Following traditional Gaussian splatting, we adaptively control the number of Gaussians during optimization. Periodically throughout the optimization, we duplicate Gaussians with high gradient magnitudes $\partial \mathcal{L}/\partial \mathbf{r}$, indicating areas where the model struggles to fit the data. We also periodically prune Gaussians with very small amplitude $q_a$ that are not contributing substantially to projection measurements and are thus unlikely to represent real atoms.

\subsection{Atomic Priors}
A key advantage of our Gaussian representation compared to volumetric representations is the ability to incorporate known physical priors about atomic structures. Unlike natural images where Gaussian splatting was originally introduced, images of atoms are highly structured in ways that are easy to express with our parameterization. We introduce two simple, easy-to-implement constraints to ensure that our reconstructions are physically plausible and there is a one-to-one correspondence between Gaussians and atoms. 

\textit{Isotropy Constraint:}
Since atoms generally exhibit spherically symmetric electron density distributions, we enforce isotropy by learning a single standard deviation $\sigma_a$ per Gaussian, and constructing a covariance matrix that is spherically symmetric with this standard deviation. In other words, we set $\Sigma_a=\sigma_aI$, where $I$ is the identity matrix.

\textit{Interatomic Distance Constraint:}
To enforce physically-plausible atomic arrangements, we periodically apply a minimum interatomic distance constraint as the optimization progresses. At each stage, we compute the pairwise distances between all Gaussians. For each Gaussian, we find the neighbors within a threshold distance, and replace them all with a single Gaussian with the mean amplitude, location, and scale of the neighbors. This process ensures that no atoms are within the threshold distance of each other in the final structure. Computing and storing pairwise distances across the entire set of atoms is computationally costly for particles with large numbers of atoms. For particles with more than 10,000 atoms, we instead search each Gaussian's top 20 nearest neighbors for candidates within the threshold distance. Since this process is repeated periodically, clusters merge gradually  over the course of the optimization, and the constraint on interatomic distances is still strongly enforced in practice.

\section{Experimental Results}
\subsection{Simulated Data}
We perform a comprehensive evaluation of our method on simulated data where the ground truth structure is known, allowing us to verify the correctness of the approach. We also demonstrate a proof-of-concept result on an experimental dataset.

We generate synthetic datasets of four nanoparticles chosen to cover a variety of sizes, atomic compositions, and amount of periodic structure:
\begin{itemize}
    \item \textbf{Au}: spherical face-centered cubic gold nanoparticle with both stacking faults and twin defects.
    \item \textbf{FePt}: iron-platinum nanoparticle with grain boundaries.
    \item \textbf{aPd}: amorphous pure palladium nanoparticle.
    \item \textbf{HEA}: amorphous nickel-palladium-platinum high-entropy alloy.

\end{itemize}
A table comparing characteristics of each particle is available in the Supplementary Material.

To simulate projection measurements, we first compute the electrostatic potential of the atomic configuration of each nanoparticle using abTEM~\cite{madsen2020abtem}. We then use the forward model in Eq.~\ref{eq:forward_model} to compute projection measurements at each tilt angle at a resolution of 0.25\r{A}. We simulate the effects using a finite-size electron probe by convolving the projected potential with a Gaussian blur kernel  with a standard deviation of 0.5\r{A}. 

Finally, we apply noise to simulate deviations from the idealized imaging process. In practice, our data are ptycho-tomographic, where a ptychography dataset is acquired and reconstructed for each tomographic angle. A ptychography dataset consists of many individual diffraction patterns that each have a Poisson noise profile. These datasets are reconstructed via an iterative algorithm into phase images used as tomographic projections, and this work focuses on the tomographic reconstruction of such these images dataset. While the raw measurement noise is Poisson, there is no clear statistical characterization of the noise after propagation through the ptychographic reconstruction algorithm. In the absence of a tractable noise model, we use additive Gaussian noise as a stand-in to represent stochastic variations from the forward model.

The physical prior imparted by our Gaussian representation should alleviate some issues with the ill-posedness of the inverse problem in Eq.~\ref{eq:voxel_objective}. To study this effect, we simulate two tomographic configurations for each sample: one with all tilt angles at 1\degree~spacing between $\pm$90\degree~and one with angles at 3\degree~spacing between $\pm$70\degree, similar to the missing wedge encountered in experimental settings. 

We compare our Gaussian splatting approach against several tomographic reconstruction techniques:

\begin{itemize}
    \item \textbf{Filtered Backprojection (FBP)} is a widely-used classical method for tomographic reconstruction, especially in computed tomography. It spreads each projection measurement back across the rays that contributed to it, high-pass filtering the projection data to preserve high-frequency details in the image. We apply standard 2D FBP to each slice along the projection axis to form our 3D FBP reconstruction.
    \item \textbf{Simultaneous Algebraic Reconstruction Technique (SART)}~\cite{andersen1984simultaneous} is an algebraic method for tomographic reconstruction that iteratively updates the image estimate based on differences between projection measurements and the forward model applied to the current image estimate. 
    \item \textbf{Implicit Neural Representations (INRs)} are multi-layer perceptrons trained via a data consistency loss to output the voxel intensities at given continuous input coordinates. We train an INR for each projection dataset as described in~\cite{chien2023space}.
\end{itemize}

\begin{figure*}[htbp]
\centering
\includegraphics[width=\textwidth]{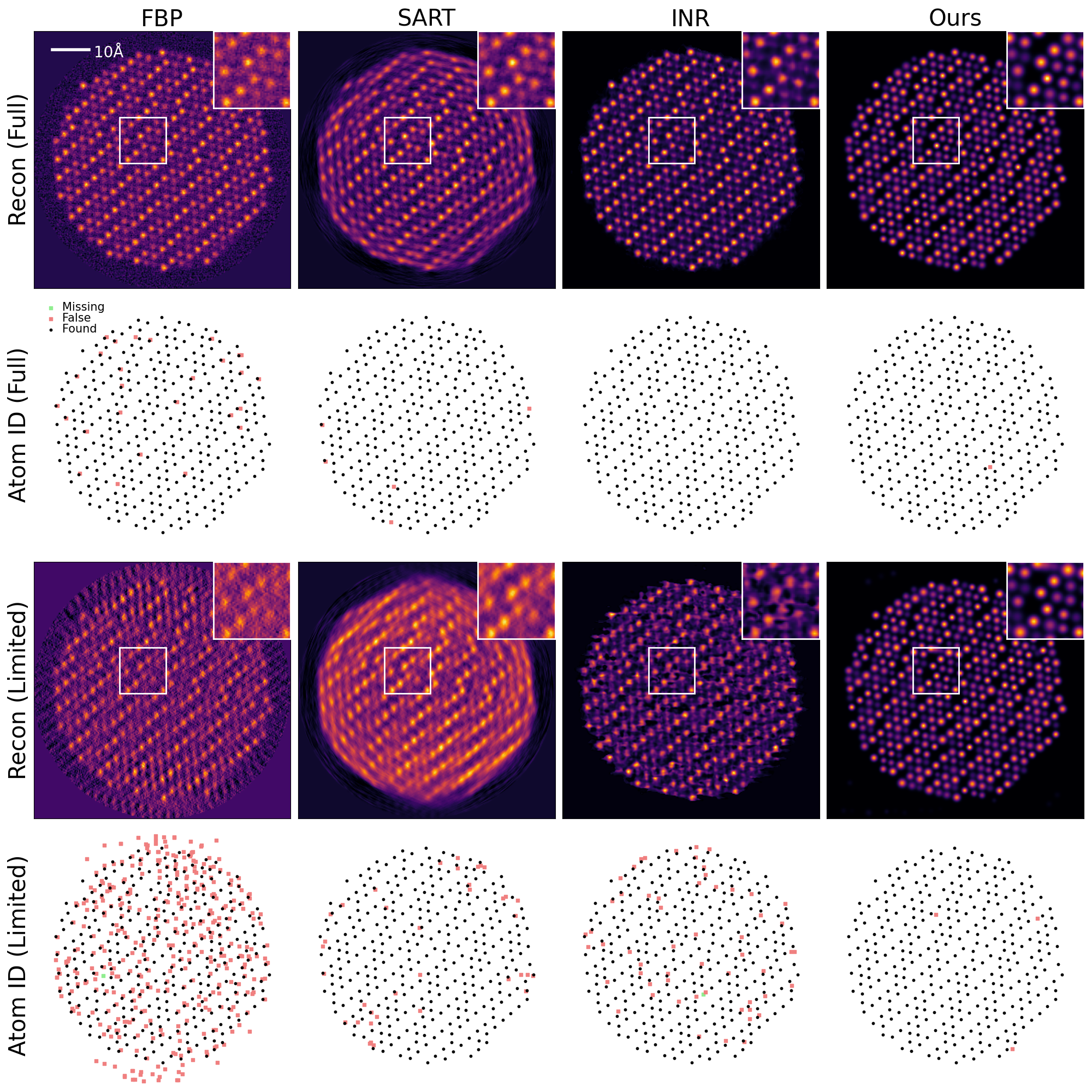}
\caption{A single (central) slice of the 3D reconstruction and its corresponding atom identification for the simulated gold nanoparticle, for both the full-data case where all projection angles are  available and the case of limited data (angles between -70\degree~and -68\degree~in 3\degree~increments). We compare our reconstruction, which directly solves for atom properties, to several baseline algorithms that use volumetric reconstruction followed by atom tracing: Filtered Backprojection (FBP), Simultaneous Algebraic Reconstruction Technique (SART), and Implicit Neural Representation (INR). All methods perform fairly well in the case with no missing wedge, but the baseline performance degrades in the missing wedge case. Our method accurately finds the majority of atoms without yielding excessive false positives. }
\label{fig:reconstruction_comparison}
\end{figure*}   
Each of these baseline reconstruction methods produces a volumetric reconstruction, which requires subsequent atom tracing to extract atomic coordinates. In each case, we run a simplified first step of atom tracing in the style of~\cite{yang2017deciphering}, where we identify all local intensity peaks, and iteratively add candidate peaks to our atom list if they are above a 2.0\r{A} threshold in distance from the closest previously-identified peak.

On simulated data where ground truth is available, we assess our method quantitatively using metrics that characterize both reconstruction quality and atomic position accuracy. Specifically, we evaluate reconstruction quality in terms of the structured similarity index metric (SSIM) between the ground truth and reconstructed atomic potentials. Our quantitative evaluation of atom identification focuses on the accurate identification of atom locations. We match each ground truth atom to the closest algorithmically-identified atom location and report the rates at which (1) atoms are found and (2) false positives are reported, considering an atom correctly identified if its position is within 0.75\r{A} of the ground truth, less than half the typical bond length for atoms in the particles we study.

Figure~\ref{fig:reconstruction_comparison} shows a qualitative comparison of the gold nanoparticle reconstructions obtained from all methods, for both the case with all projection measurements and a dataset affected by the missing wedge. All methods recover accurate reconstructions in the ideal case, and the atom tracing procedure is largely accurate. In the missing wedge case, however, the baseline methods suffer from reconstruction artifacts that yield inaccurate atom identification. Noise in the filtered backprojection reconstruction, blurring in the SART reconstruction, and streaking in the INR yield either missed or artificially identified local maxima.  Our Gaussian-based approach, which directly solves for the atom locations instead, largely avoids these errors. The atoms learned by our method are spherical and appropriately spaced as enforced by our parameterization and physical priors, while the baseline methods reconstruct physically unrealistic structures that lead to downstream atom tracing mistakes. Qualitative results for all other particles in our dataset are in the Supplementary Material.

Quantitatively, Figure~\ref{fig:quantitative_results} shows this trend holds across our dataset of simulated particles. In general, our method performs comparably to the baselines in the setting of full data availability. In the limited data setting, our method outperforms the baselines in terms of reconstruction quality and false positive rate while maintaining a similar true positive rate. The Supplementary Material contains results of an ablation study further describing the performance of all methods in more specific cases of the limited data setting, isolating the effects of angle spacing from a contiguous missing wedge.

\begin{figure}[h]
    \centering
    \includegraphics[width=\columnwidth]{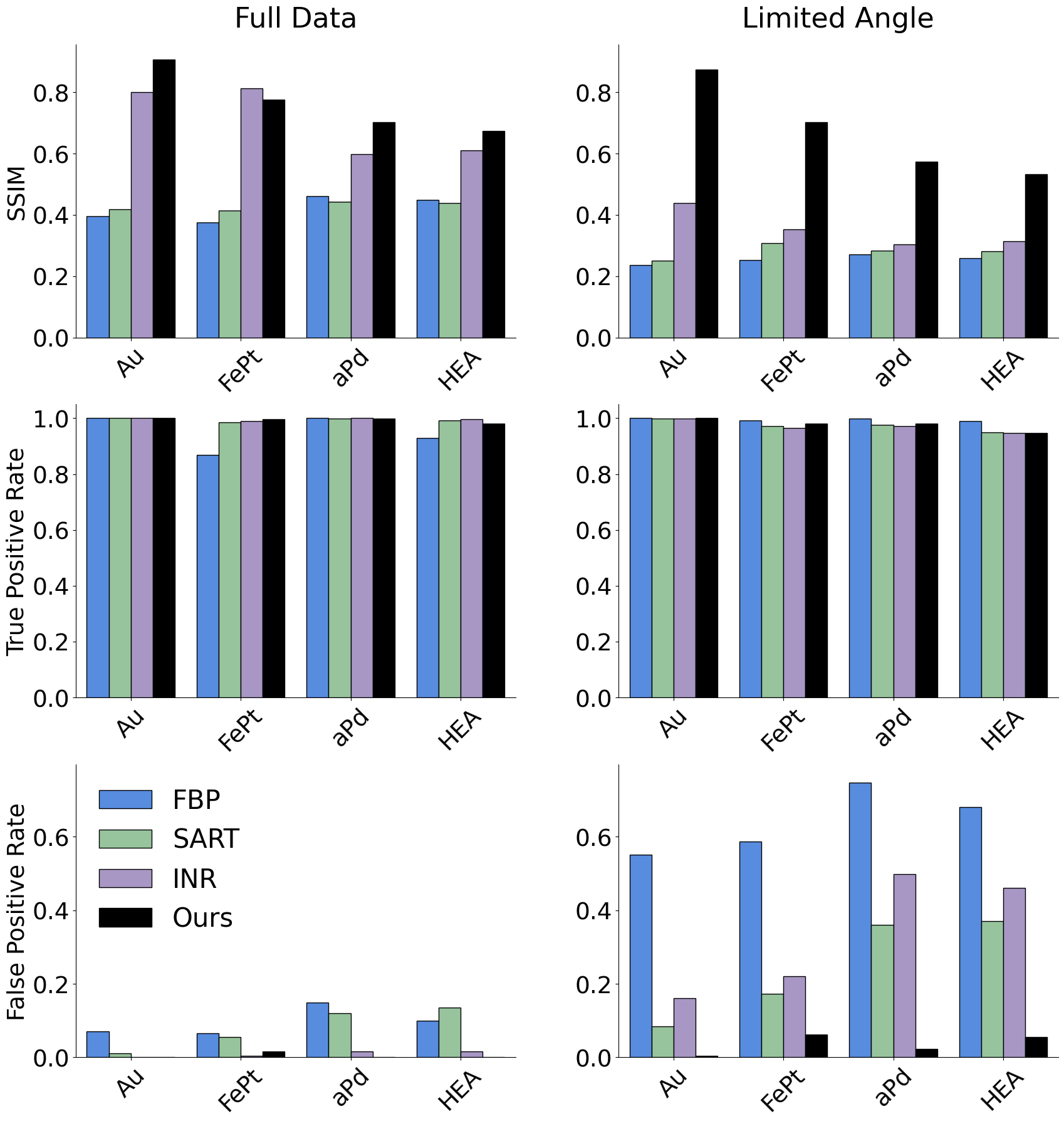}
    \caption{Quantitative results on the particles in our dataset. All methods achieve a fairly high true positive rate on data with all angles, and this performance degrades slightly in the limited angle setting. However, other methods' performance degrades dramatically in terms of false positive rate in the limited data setting, while our method is less affected. Similarly, the reconstruction quality of other methods decreases more dramatically in the limited data setting.}
    \label{fig:quantitative_results}
\end{figure}

\begin{figure*}[h]
    \centering
    \includegraphics[width=\textwidth]{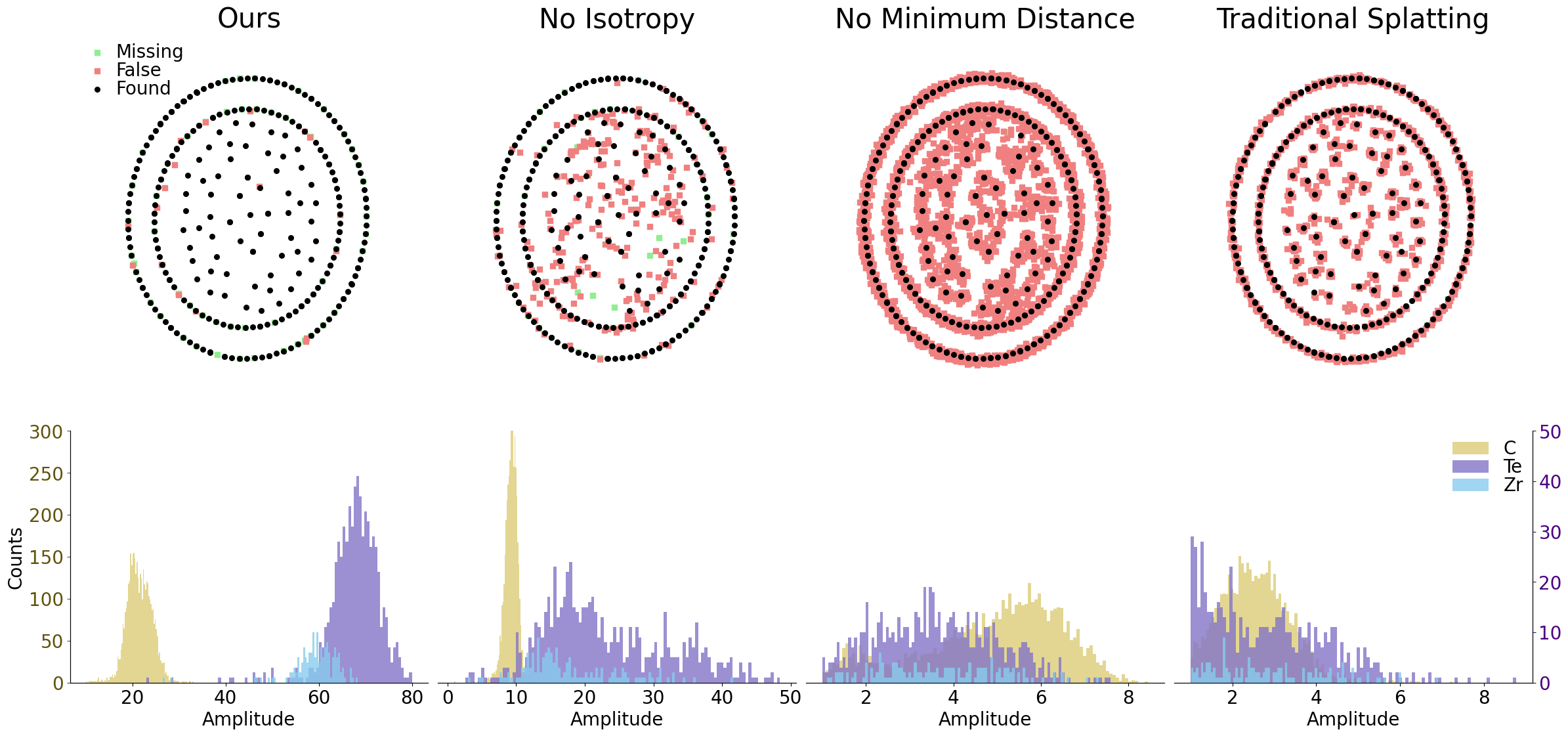} %
    \caption{Ablation study showing the effect of physical constraints on our method's structure identification. Top row: atom localization for an idealized slice of the simulated double-wall Zirconium-Tellurium carbon nanotube with all projection angles and no noise or probe blur. Bottom row: learned Gaussian amplitude for matched atoms. When the isotropic and minimum distance constraints are not included, multiple Gaussians are fit to each atom, yielding an inaccurate structural representation. Each Gaussian is typically fit with a lower amplitude to compensate for the presence of several other atoms close by (note the different x-axis scales). Our physical constraints force a 1-to-1 correspondence between Gaussians and atoms, which in turn allows their learned opacities to distribute separably by atomic species.}
    \label{fig:ablation}
\end{figure*}

Next, we perform an ablation study to investigate the importance of the physical priors we enforce to adapt traditional Gaussian splatting specifically for our atomic problem. We run this experiment on a simulated double-wall Zinc-Tellurium carbon nanotube with no additional artifacts (i.e. no missing wedge, blur, or Gaussian noise). Figure~\ref{fig:ablation} shows the effect of removing each physical prior on the atom correspondence. 
Each method correctly finds nearly all of the atoms in the sample. However, the ablated versions of our method without physical constraints tend to use multiple Gaussians to represent each atom. While these solutions may yield projections that fit the data equally well, they are not sufficient for our problem of atomic structure identification. We also show that the physical constraints improve separation of different atomic species by their learned intensity. The intensity distributions for carbon, tellurium, and zirconium atoms are much more separable when the physical constraints are enforced, and follow the ordering expected based on atomic number. Only the reconstruction that incorporates the full set of physical priors achieves the 1-to-1 Gaussian to atom correspondence that is necessary for atomic structure identification.

\subsection{Experimental Data}
For preliminary experimental validation, we use experimentally collected projection data of a Zr-Te sandwich carbon nanotube from the National Center for Electron Microscopy (NCEM) at Lawrence Berkeley National Laboratory (LBNL). This data was acquired with atomic electron ptychotomography, and the ptychographic reconstructions were manually aligned and averaged across several sections of the periodic nanotube to achieve higher SNR. These data were previously published and analyzed in~\cite{pelz2023solving}.

\begin{figure*}[htbp]
\centering
\includegraphics[width=\textwidth]{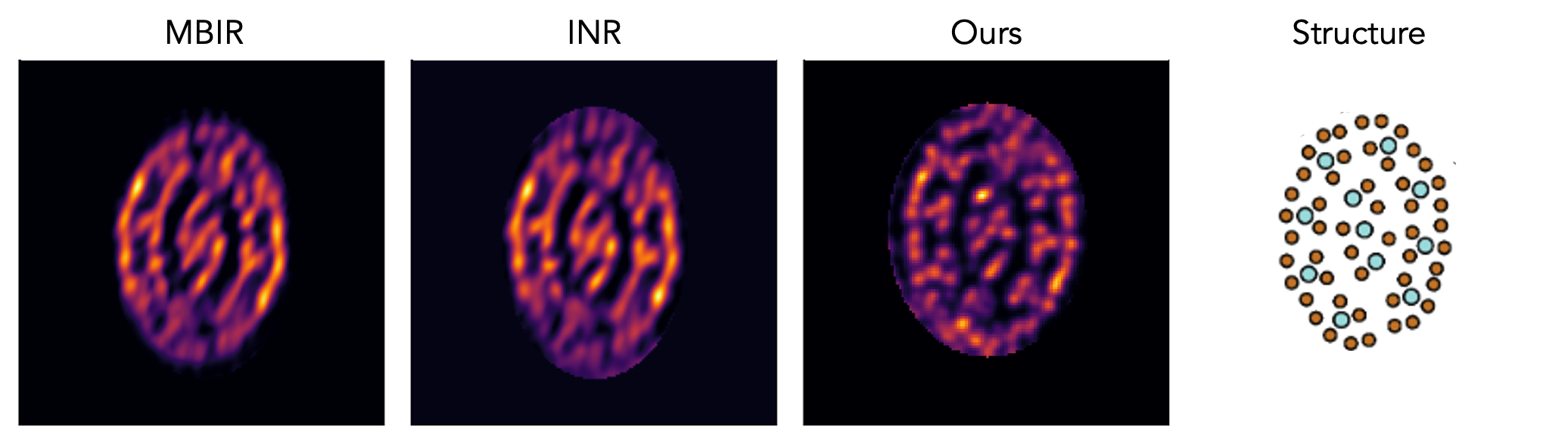}
\caption{Projection along a single repeating unit of a double-wall Zirconium-Tellurium nanotube reconstructed from experimental data. Our reconstruction is compared with a previously published~\cite{pelz2023solving} model-based image reconstruction (MBIR) and the output of an implicit neural representation (INR). All reconstructions suffer from significant artifacts, but our reconstruction preserves some atomic structure that is harder to observe in the baselines. In particular, individual atoms are clearly discernible at the center of the tube in an arrangement mimicking the known structure of the particle (reproduced from~\cite{pelz2023solving}).}
\label{fig:experimental_results}
\end{figure*}
Figure~\ref{fig:experimental_results} shows the reconstruction of the Zr-Te double-walled nanotube from experimental projection data. For clarity, we omit the tube edges, which are artifacted in all reconstructions  because of a missing wedge artifact in the horizontal dimension and the presence of closely-spaced light Carbon atoms on the tube walls that are difficult to distinguish. We focus instead on the heavy atoms at the center of the carbon nanotube. Compared to a previously published model-based reconstruction and an INR baseline, our reconstruction provides a clearer reconstruction that matches the solved structure of the particle. In particular, individual columns of atoms are visible in the center of the structure that are not separated in the baseline reconstructions. Our reconstruction is promising as a proof of concept that a Gaussian parameterization could be effective in recovering heavy atom structures from real-world data.

\section{Discussion}
At the core of our contribution is the reformulation of the reconstruction task as an optimization with respect to atomic parameters rather than voxel values, which has significant implications for the conditioning of the inverse problem. In standard volumetric reconstruction, the inverse problem solves for voxel intensities, a problem well-studied in classical tomography literature, with the condition number typically scaling with the resolution of the reconstruction and the completeness of the projection data. The number of unknowns is the number of voxels to be reconstructed, while the number of measurements is proportional to the number of projection angles. In contrast, our approach directly solves for atomic parameters (positions, scales, and intensities). A typical particle that we study contains several thousand atoms, so in theory this approach improves the conditioning of the inverse problem relative to the case of solving for network weights or individual voxel values, of which there may be several hundreds of thousands or millions. However, the variable number of Gaussians makes it more difficult to reason about the conditioning of the problem. In particular, any individual Gaussian could be replaced with a sum of many smaller Gaussians that provide the same total intensity contribution in any projection direction, creating a degenerate solution space. Further, particularly on experimental data, small deviations from idealized data can be easily overfit by adding extra Gaussians to fit these nuisance variations. Our minimum distance constraint enforced via clustering is thus key for correctly solving the problem; it effectively reduces the total number of Gaussians that can be solved for, thereby preventing the inverse problem from exploding in the number of variables to be optimized. 

Two practical limitations of AET also significantly affect the conditioning of the inverse problems. The first is the missing wedge problem; the tilt range of our experiments is typically restricted to approximately $\pm$70\degree~from the horizontal axis. This results in a wedge-shaped region of unsampled information in Fourier space, leading to elongation artifacts and reduced resolution in conventional reconstructions. The missing projection measurements make the inverse problem ill-posed, and reconstructions with these elongation artifacts are a solution that is consistent with the acquired data. Our Gaussian splatting approach effectively imparts a strong enough prior to disambiguate the solutions; only the solution with spherical atoms fits the Gaussian parameterization.

The second practical limitation is resolution. Beam-induced blur and other imaging artifacts can make it difficult to distinguish neighboring atoms. This does not affect the conditioning of the inverse problem with respect to voxel values; even if we cannot distinguish which atom contributed to a particular voxel, we can solve for that voxel's intensity during the volume optimization stage. However, there is then no guarantee that the subsequent atom tracing step will succeed, as it is likely that adjacent atoms whose intensities touch after the blur will be lumped together as one atom.  Our method directly incorporates atomic priors which may help during the optimization process. If atoms are overlapping but not entirely merged, it is possible to fit two Gaussians to each peak under our parameterization. That said, if the blurred atoms are so close as to be indistinguishable, no method will be able to resolve them individually.

\subsection{Limitations}
While our results show promise in simulation and experiment, we acknowledge several limitations of our proposed method.
First, our evaluation focuses largely on simulated datasets, and comprehensive validation on diverse experimental data remains ongoing. 
In particular, our simulation strategy introduces a Gaussian blur kernel, and the similarity between the structure of this kernel and the Gaussian parameterization of our model may raise suspicion of an \textit{inverse crime}, where the same forward model is used to simulate synthetic data and to algorithmically reconstruct that data. We note that our Gaussian simulation is based in known physics of the imaging system. Atomic densities are modeled with Lobato potentials~\cite{lobato2014accurate}, which are then convolved with the point spread function of the microscope, commonly modeled as Gaussian. The resulting data that we fit is not exactly Gaussian but approximately so, which also appears true in experiment.
Our proof-of-concept result on experimental data, where we have no control over the data generation process but can see the core atomic structure, provides promising evidence that this is the case.

Nevertheless, our simulation procedure does not model several potential artifacts in the AET acquisition process, and additional work is needed to ensure the method works as an out-of-the-box tool for AET practitioners studying arbitrary samples. For example, we do not explicitly model carbon contamination, a common issue in electron microscopy experiments~\cite{griffiths2010quantification}. Carbon deposits can accumulate on the sample during imaging, affecting projection quality and potentially introducing artifacts in the reconstruction. Future work will incorporate models for background contamination~\cite{chien2023space} to address this limitation.

Further, the current method assumes perfect alignment of projection images with each other. In practice, experimental data contains alignment errors that must be corrected during reconstruction. While our physical priors may provide some robustness to misalignment, explicit treatment of alignment uncertainty within the Gaussian splatting framework represents an important direction for future research.

\section{Conclusion}

We have presented a novel approach to atomic electron tomography that uses a Gaussian parameterization with physics-based priors to directly reconstruct atomic structures from projection data. By reformulating the reconstruction problem in terms of atomic parameters rather than voxel values, our method achieves physically plausible reconstructions and accurate atomic structures that are more robust to imaging artifacts than other techniques.

The incorporation of our physical constraints significantly improves the conditioning of the inverse problem and enables reliable learning of 1-to-1 Gaussian to atom correspondences even in challenging imaging conditions. Simulation and experimental validation confirm our method's potential for resolving a wide variety of atomic structures without making strong assumptions (e.g. periodicity of the sample).

Our approach has the potential to accelerate materials characterization and discovery across numerous scientific disciplines, from catalyst design to nanostructure engineering. Future work will focus on extending the approach to more complex particles and corruption classes, and addressing the limitations identified in this study. Beyond this, Gaussian-based representations show promise for inverse problems in \textit{any} type of atomic imaging -- for example, 2D ptychographic imaging. 

\ifpeerreview \else
\section*{Acknowledgments}
This work was supported by the U.S. Air Force Office Multidisciplinary University Research Initiative (MURI) program under award no. FA9550-23-1-0281 and by STROBE: A National Science Foundation Science and Technology Center under Grant No. DMR 1548924. Compute resources were provided from an NVIDIA Academic Grant. Laura Waller is a Chan Zuckerberg Biohub SF investigator.
\fi

\bibliographystyle{IEEEtran}
\bibliography{references}

\onecolumn

\section{Supplementary Material}  
\subsection{Implementation Details}
We implement our method using the nerfstudio~\cite{tancik2023nerfstudio} software package which in turn uses the gsplat~\cite{ye2025gsplat} Gaussian splatting implementation, though we adapt the Gaussian splatting backward and forward passes for our tomographic imaging model.

We scale our input data to the range [0,256] and scale the physical space of the volume to the range[-0.5,0.5] for optimization and then rescale back to physical parameters once the optimization is complete. This scaling allows us to remain consistent with traditional Gaussian splatting defaults and therefore benefit from their default hyperparameter choices. These choices largely work for our problem, with one exception: we find that our method is highly sensitive to the threshold for densifying Gaussians, which we set to be 0.005 for all experiments. Our method is also somewhat sensitive to the clustering threshold distance, which we set to be equivalent to approximately 0.25\r{A}. This is smaller than a typical inter-atomic bond distance, but we find that larger settings for this parameter prevent smooth optimization as Gaussians are not able to pass by other Gaussians to new locations. In practice, this is sufficient to achieve the 1-to-1 Gaussian to atom correspondence desired for our application.

We initialize the method with 10,000 randomly placed Gaussians and run the optimization for 10,000 iterations. Our method takes $\sim$7 minutes to run on an NVIDIA A6000 GPU and requires up to 5 GB of GPU memory for the largest particle we studied.

\subsection{Dataset}
\begin{table}[h]
\centering
\begin{tabular}{lccc}
\hline
\textbf{Particle} & \textbf{\# Atoms} & \textbf{Atomic Species} & \textbf{Structure} \\
\hline
Au      & 6,679  & Au & Periodic \\
aPd     & 16,755   & Pd  & Amorphous \\
FePt    & 8,344   & Fe, Pt  & Periodic \\
HEA     & 16,750 & Ni, Pd, Pt   & Amorphous \\
\hline
\end{tabular}
\vspace{3pt}
\caption{Summary of particle characteristics including number of atoms, species composition, and periodicity of the structures in our simulation dataset.}
\label{tab:particle_summary}
\end{table}

\subsection{Missing Wedge Ablation}
\begin{figure}[h]
    \centering
    \includegraphics[width=\linewidth]{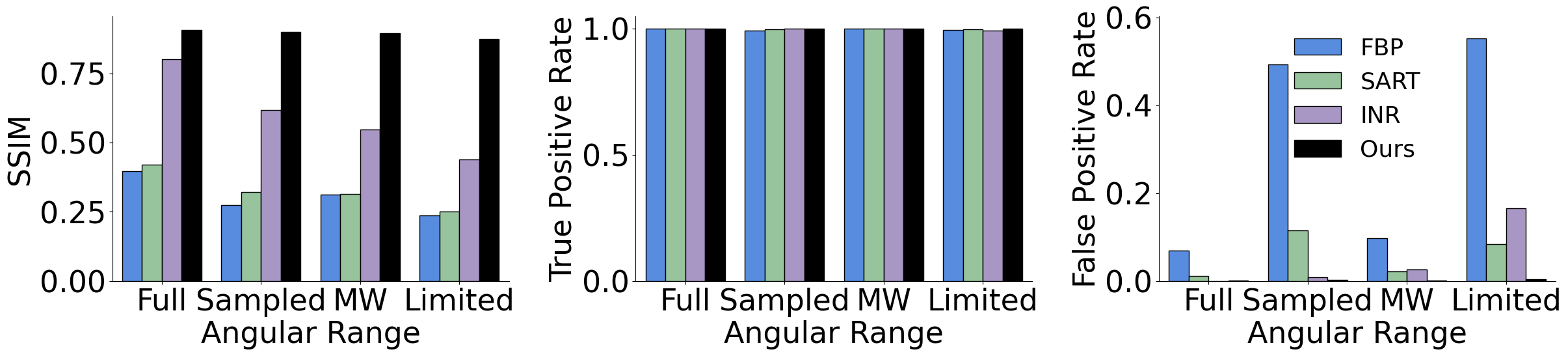}
    \caption{Ablation study isolating the effects of angular sampling sparsity (“Sampled”) and the contiguous missing wedge (“MW”) on Au particle reconstruction quality. These settings are compared to the full data setting ("Full") and the setting reported in the main manuscript where both angular sampling sparsity and a contiguous missing wedge are present ("Limited"). We compare SSIM, true positive rate, and false positive rate for four methods: FBP, SART, INR, and Ours. The performance of FBP and SART degrade more in the Sampled setting than the MW setting, while the opposite is true is true for the INR. Our method consistently outperforms the baseline methods across all settings.}

    \label{fig:mw_ablation}
\end{figure}

\newpage

\subsection{Additional Particle Reconstructions}
\begin{figure*}[h]
\centering
\includegraphics[width=\textwidth]{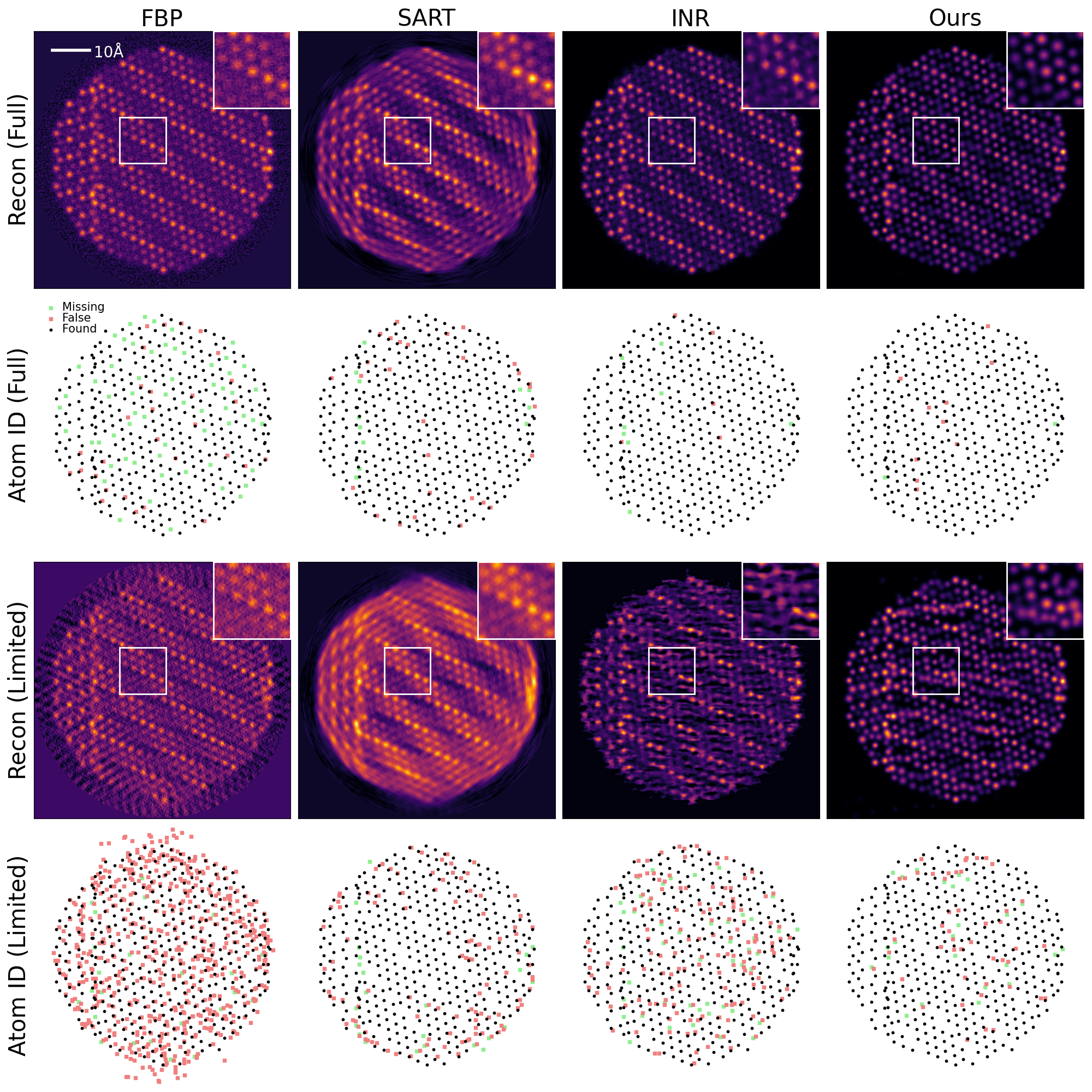}
\caption{A single (central) slice of the 3D reconstruction and its corresponding atom identification for the simulated iron-platinum nanoparticle, for both the full-data case where all projection angles are  available and the case of limited data (angles between -70\degree~and -68\degree~in 3\degree~increments). We compare our reconstruction, which directly solves for atom properties, to several baseline algorithms that use volumetric reconstruction followed by atom tracing: Filtered Backprojection (FBP), Simultaneous Algebraic Reconstruction Technique (SART), and Implicit Neural Reconstruction (INR). All methods perform fairly well in the case with no missing wedge, but the baseline performance degrades in the missing wedge case. Our method accurately finds the majority of atoms without yielding excessive false positives. }
\label{fig:reconstruction_comparison_fept}
\end{figure*}  

\begin{figure*}[htbp]
\centering
\includegraphics[width=\textwidth]{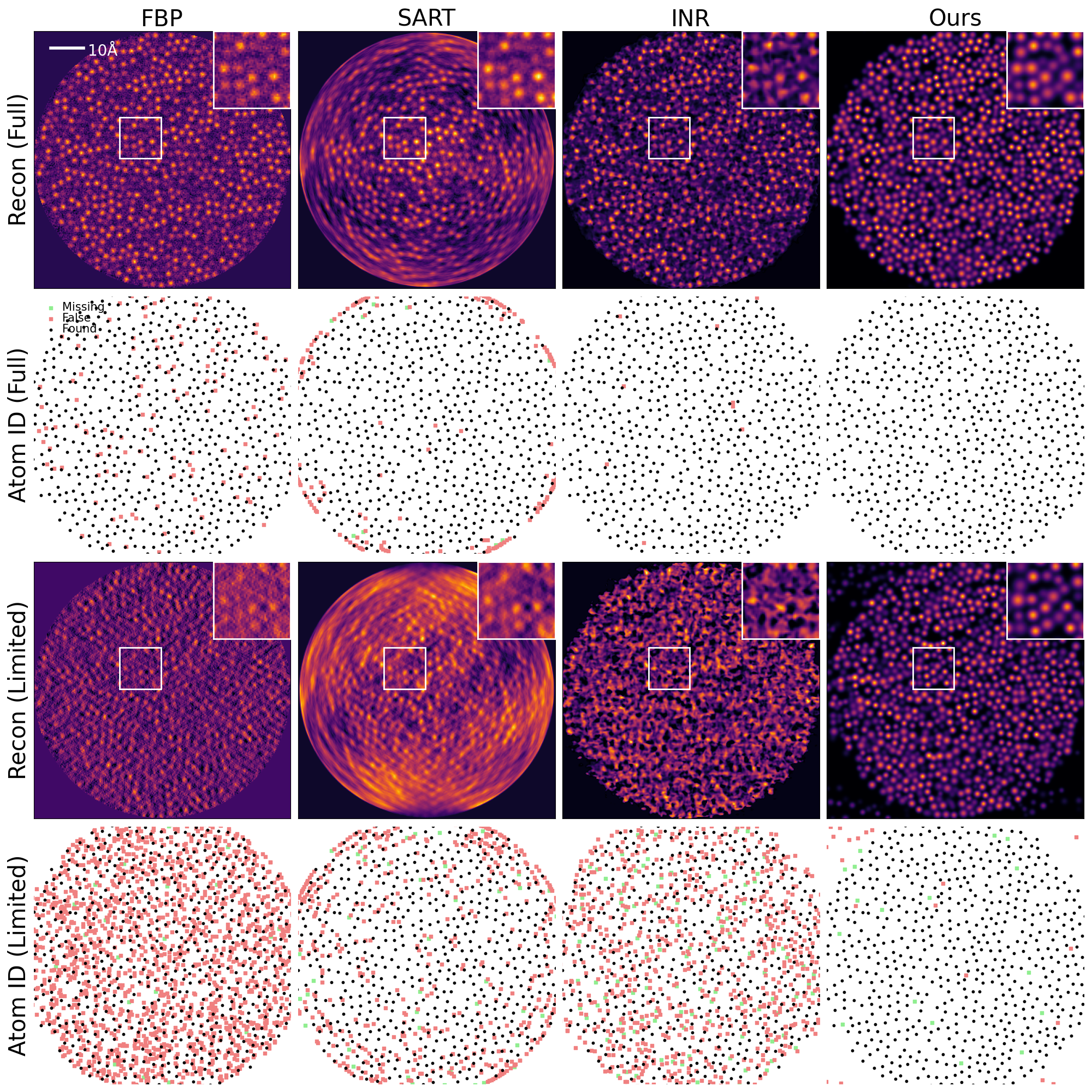}
\caption{A single (central) slice of the 3D reconstruction and its corresponding atom identification for the simulated amorphous Palladium nanoparticle, for both the full-data case where all projection angles are  available and the case of limited data (angles between -70\degree~and -68\degree~in 3\degree~increments). We compare our reconstruction, which directly solves for atom properties, to several baseline algorithms that use volumetric reconstruction followed by atom tracing: Filtered Backprojection (FBP), Simultaneous Algebraic Reconstruction Technique (SART), and Implicit Neural Reconstruction (INR). All methods perform fairly well in the case with no missing wedge, but the baseline performance degrades in the missing wedge case. Our method accurately finds the majority of atoms without yielding excessive false positives. }
\label{fig:reconstruction_comparison_apd}
\end{figure*}  

\begin{figure*}[htbp]
\centering
\includegraphics[width=\textwidth]{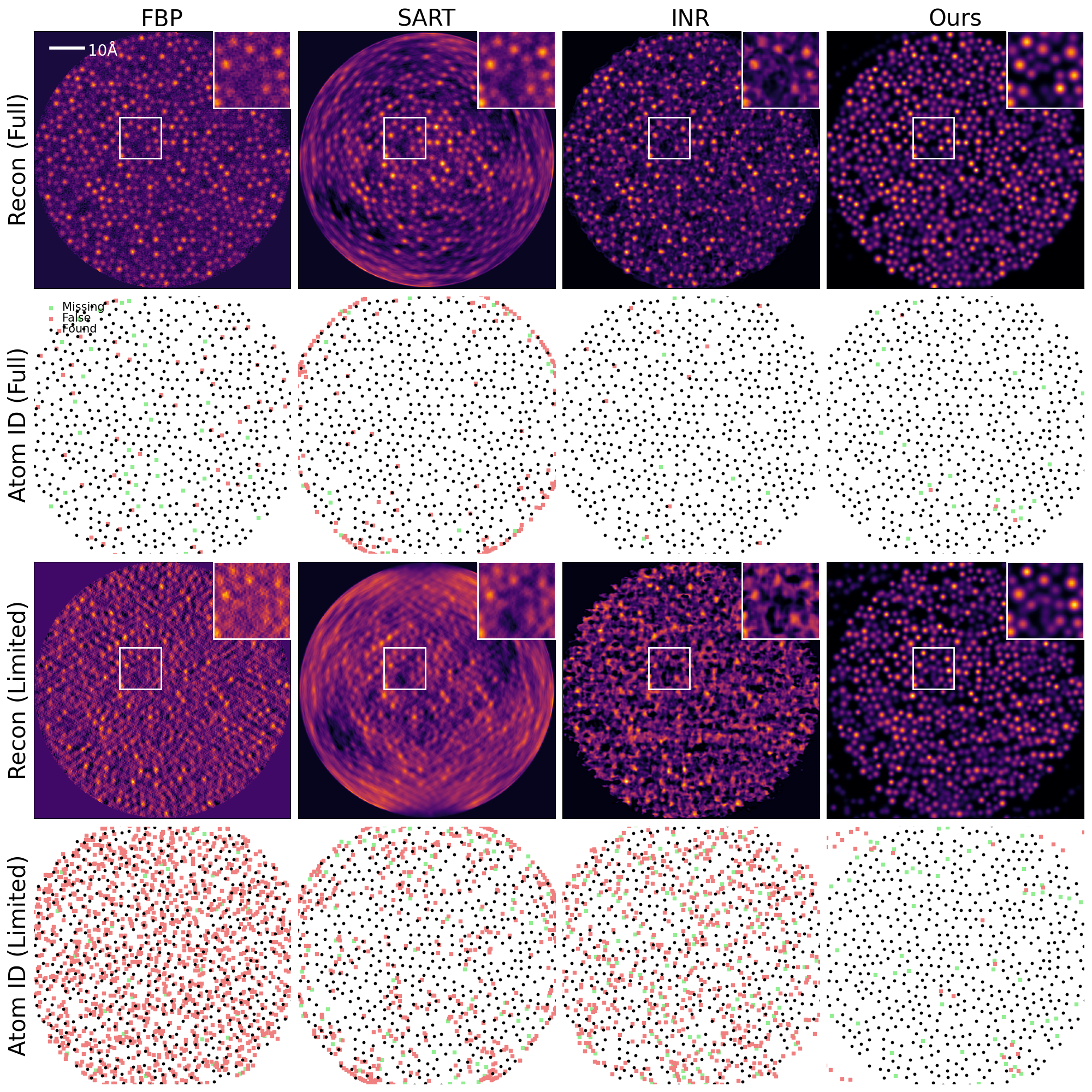}
\caption{A single (central) slice of the 3D reconstruction and its corresponding atom identification for the simulated high-entropy alloy nanoparticle, for both the full-data case where all projection angles are  available and the case of limited data (angles between -70\degree~and -68\degree~in 3\degree~increments). We compare our reconstruction, which directly solves for atom properties, to several baseline algorithms that use volumetric reconstruction followed by atom tracing: Filtered Backprojection (FBP), Simultaneous Algebraic Reconstruction Technique (SART), and Implicit Neural Reconstruction (INR). All methods perform fairly well in the case with no missing wedge, but the baseline performance degrades in the missing wedge case. Our method accurately finds the majority of atoms without yielding excessive false positives. }
\label{fig:reconstruction_comparison_hea}
\end{figure*}  

\ifpeerreview \else

\fi

\end{document}